\documentstyle[preprint,aps]{revtex}

\preprint{\bf PREPRINT}
\input psfig

\begin{document}
\columnsep0.1truecm
\draft
\title{Acoustic emission from crumpling paper}
\author{Paul A. Houle and James P. Sethna}
\address{Physics Department, Cornell University, Ithaca, NY, 14853-2501}
\maketitle

\begin{abstract}

	From magnetic systems to the crust of the earth,  many
physical
systems that exhibit a multiplicity of metastable states emit pulses
with a broad power law distribution in energy.  Digital audio
recordings
reveal that paper being crumpled,  a system that can be easily held
in
hand,  is such a system.
Crumpling paper both using the traditional hand method and a novel
cylindrical geometry uncovered a power law distribution of pulse
energies
spanning at least two decades: $p(E)=E^\alpha$, $\alpha=1.3-1.6$.
Crumpling initially flat sheets into a
compact ball (strong crumpling),  we found little or no
evidence that the energy
distribution varied systematically over time or the size of
the sheet.  When we applied repetitive small deformations
(weak crumpling)
to sheets which had been previously folded along a regular grid, 
we found
no systematic dependence on the grid spacing.  Our results
suggest that the
pulse energy depends only weakly on the size of paper regions
responsible
for sound production.

\end{abstract}


\narrowtext

	Everyone has had the experience of crumpling an unwanted sheet
of paper into a compact ball prior to disposing of it.  In one trial,
after we hand crumpled a sheet of Xerox 4024 paper with an area of
approximately 600 ${\rm cm}^2$ and thickness .1 mm ${\rm cm}^3$
produced a resulting metastable object that appeared to be a roughly
spherical ball with a volume of about 65 ${\rm cm}^3$.  Although the
crumpled object remains a sheet, a human observer with poor vision
would perceive it to be a three dimensional object; past experiments
have shown that crumpled balls of paper and similar materials have a
fractal structure with dimension $D=2.3-2.5$. \cite{fractalA}
\cite{fractalB} Although paper is one of the most ubiquitous composite
materials, much is still unknown about the physics of paper.  Recent
research has addressed the tearing of paper \cite{tearingA}
\cite{tearingB} \cite{tearingC} as well as the friction between two
paper sheets. \cite{friction}

	Like a thin elastic sheet, paper tends to bend much more
easily than it stretches.  If one applies a slight stress to a paper
sheet it will deform into a shape with zero Gaussian curvature almost
everywhere, a developable surface\cite{curvature}; the shape of a
M\"{o}bius band, a paper strip with unusual boundary conditions, has
recently been so modeled.  \cite{band} Unlike an elastic sheet, paper
forms permanent creases under extreme local stress.  When a crumpled
ball is flattened out, a network of creases formed by crumpling is
revealed.  We refer to the polygonal regions of the sheet bounded by
the creases as {\it facets}.  Examining an unfolded crumpled sheet,
one finds that the areas of the facets vary greatly and that a sheet
with a crease network can be easily deformed into many different
metastable states ranging from nearly flat to compact balls.

	Because paper makes an audible sound while being crumpled, we
decided to probe the dynamics of crumpling by studying the sound
produced.  Crumpling sheets of several varieties of paper and similar
materials such as plastic transparencies, we discovered that most of
the acoustic power is emitted in the form of discrete pulses or
acoustic emissions (AE).

	Acoustic emissions are a versatile probe in science and
engineering.  Ultrasonic AEs provide insight into the dynamics of
materials under both mechanical \cite{microfrac} and thermal
\cite{martAE} stress.  AEs produced by the crust of the Earth are an
important probe for geologists; the largest and rarest AE events of
the Earths crust radiate energy in excess of $4 \times 10^{12}$ J
endangering people and property on the surface of the Earth.  AE is
particularly useful for the non-destructive testing of composite
materials and have already been used to study the tearing of paper
\cite{tearingA};  all fifty states now require AE safety inspection of 
fiberglass cherry picker arms. \cite{cherrypickers}  AE has been used to
study avalanches of glass beads \cite{avalanches}.  The dynamics of
magnetic systems produce (inaudible) Barkhausen
noise, a pulsed magnetic signal with properties similar to AEs.
\cite{barkA} \cite{barkB}

	The folding states of a (nearly) inextensible sheet such as
paper can be related to many other physical systems.  Recently,
connections have been drawn between the possible states of twinned
martensites and the possible foldings of a sheet.
\cite{martlinkA} \cite{martlinkB} 
In addition, ``spin origami'' mappings have been made between the
 minimum-energy states of the classical Heisenberg antiferromagnetic
 Kag\'{o}me lattice and foldings of an inextensible
 sheet. \cite{origamiA} \cite{origamiB} Crumpling and folding
 transitions in equilibrium tethered membranes have also been a
 subject of recent interest in fields ranging from biophysics to
 superstring theory.
\cite{membranesA} \cite{membranesB}

\begin{figure}
\centerline{\psfig{figure=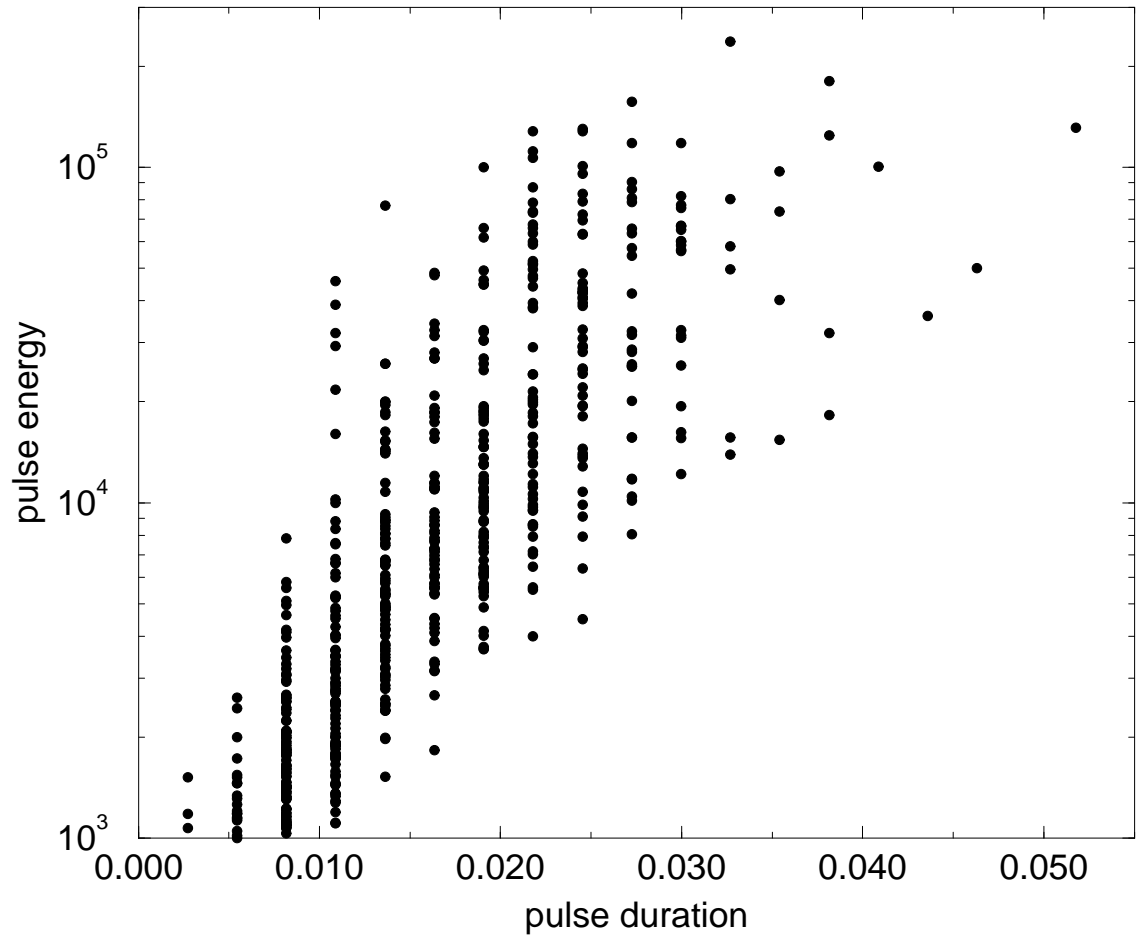,width=7.0truein}}
\caption{Scatter plot of pulse duration versus pulse energy for
cylindrical weak crumpling of Xerox 4024 paper with
a 2'' triangular
grid.  Horizontal axis is linear,  vertical axis is logarithmic.}
\label{fig:duration}
\end{figure}

\begin{figure}
\centerline{\psfig{figure=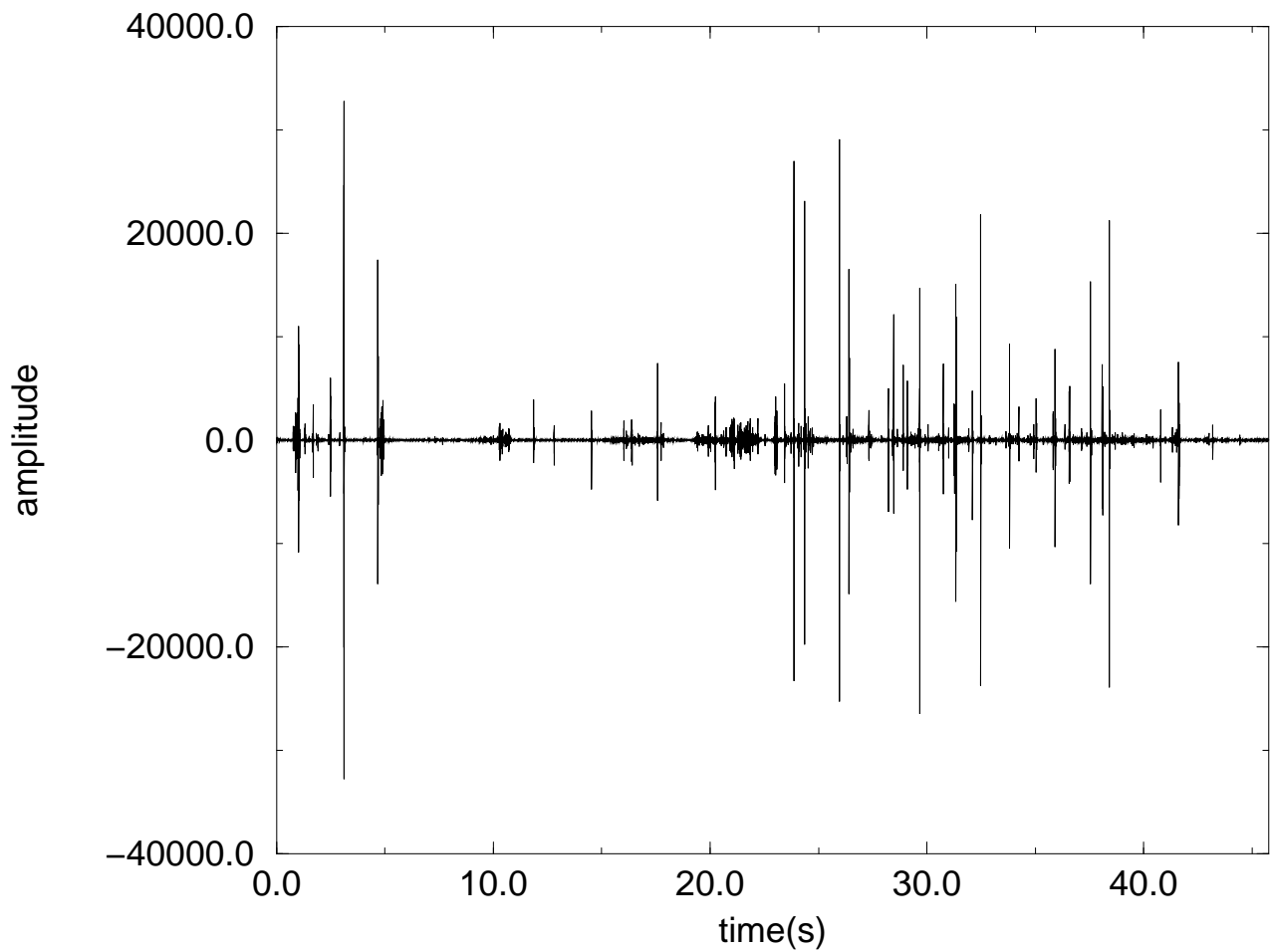,width=7.0truein}}
\caption{Sound amplitude versus time: one entire strong 
cylindrical crumple, Xerox 4024
paper 8.5'' square}
\label{fig:long}
\end{figure}

\begin{figure}
\centerline{\psfig{figure=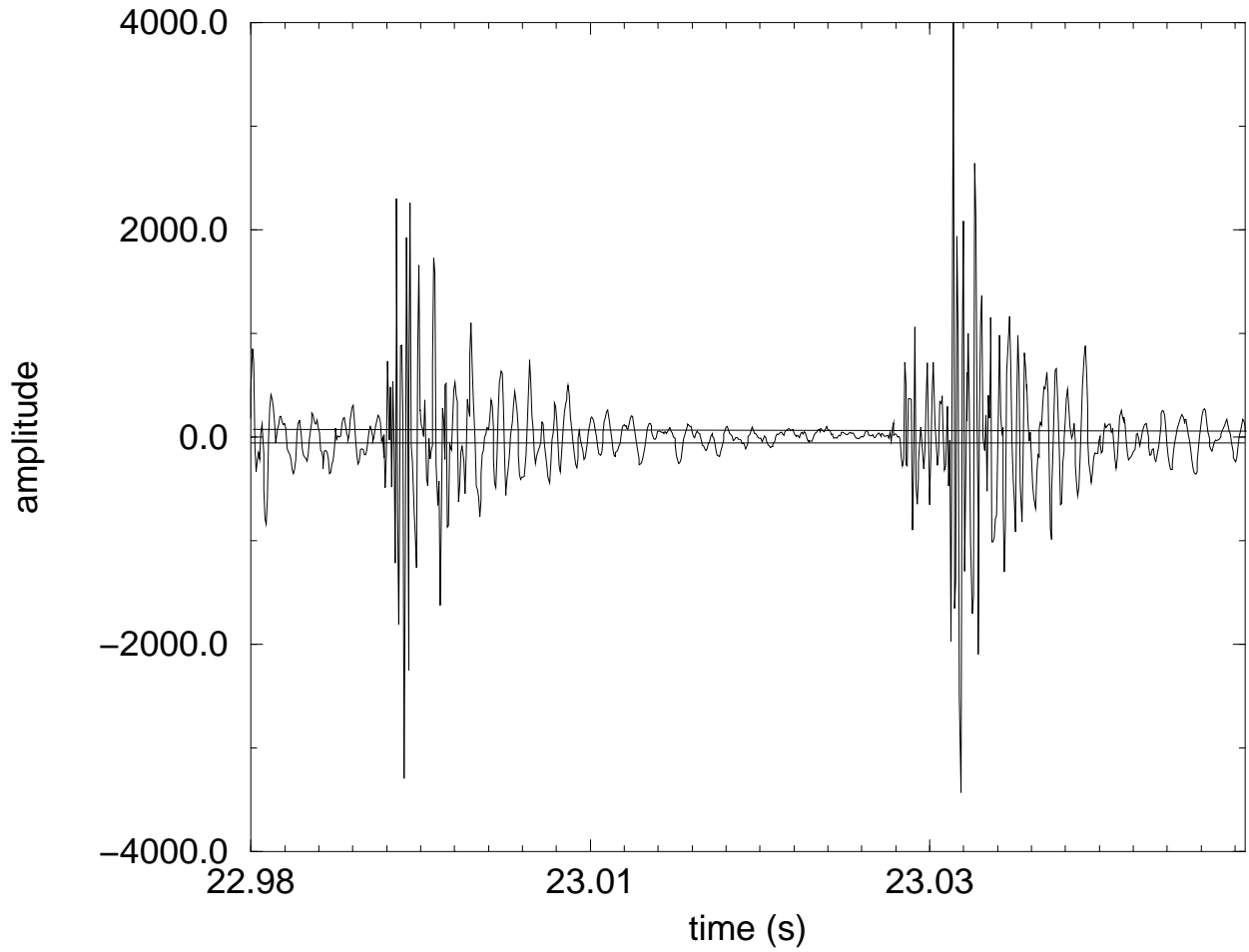,width=7.0truein}}
\caption{Sound amplitude versus time: two adjacent pulses
identified by our
algorithm.  The spacing of minor ticks is equal 
to the time bin duration in which energy
was integrated,  and the two superimposed lines show the
threshold value.  
Bins were considered ``active'' when the energy inside
equaled the bin length
times the threshold amplitude squared.}
\label{fig:short}
\end{figure}

	Crumpling paper produces pulsed AEs when facets suddenly
buckle from one configuration to another; this can be verified by
crumpling a sheet, uncrumpling it , and then slowly applying stress to
the edges by hand.  We observe that every discrete pop one hears can
be traced to a single facet of the sheet undergoing a change of
configuration; sounds do not appear to be produced directly by the
formation of creases.  Although it seems that several vibrational
modes may be excited, both the oscillation frequency, on the order of
a kilohertz, and the damping time, on the order of a millisecond,
depend strongly on the type of paper but not on the energy of the
pulse or the size of the sheet (see Figure \ref{fig:duration}).  Figure
\ref{fig:long} is the complete acoustic record of one crumpling and
Figure \ref{fig:short} shows two individual pulses separated by our
counting algorithm.  Amplitude is measured in the arbitrary units used
by the computer, where sound amplitudes are represented as signed
16-bit integers varying from $-2^{15} $ to $2^{15}-1$.

	In our experiments we used three methods to crumple paper.  In
one, {\it hand crumpling} the paper was crumpled by hand into a tight
ball as slowly and evenly as possible over a duration varying from 63
s to 74 s.  Initially it took us about 6 seconds to crumple a sheet in
hand, but we found that it was essential to crumple very slowly for
the computer to be able to identify individual events.  Hand crumpling
is interesting because it produces a very compact object, but it has
the major disadvantage that it is imprecisely defined and
irreproducible.  Particularly, hand crumpling introduces an
uncontrolled length scale related to the size of the crumpler's hands
and fingers.  Our other two methods involve fixing the paper to the
ends of two hollow cylinders using adhesive tape, then rotating the
cylinders in opposite directions by hand.  In all of our cylindrical
experiments the paper sheet was a square with sides slightly shorter
than the circumference of the cylinders, although other aspect ratios
would have been possible.  In the case of {\it strong cylindrical
crumpling} we rotated the cylinders until it was impossible to rotate
them further producing a crumpled object not quite as compact as that
produced by hand crumpling.  We also performed {\it weak crumpling}
\footnotetext[1]{suggested by Eric Kramer, private communication: see
\cite{kramerExp}} \footnotemark[1] experiments in the cylindrical
geometry, rotating the cylinders only slightly back and forth -- the
range of rotation ending just before the free edges of the sheet were
about to touch.  Cylindrical crumpling has many advantages over hand
crumpling: cylindrical crumpling can easily be performed slowly and
can be scaled precisely in size.  Because cylindrical crumpling can
crumple a sheet by applying a well-defined strain to only the edges of
the sheet, it can obviously be mechanized and may be easier to
simulate and study theoretically.  Weak crumpling, in addition, nearly
eliminates noise from friction between paper surfaces
\footnotemark[1]
and between
the paper and the hands of the crumpler.  

        We recorded audio in an anechoic chamber using a Realistic
33-1090B Pressure Zone Microphone, and a Realistic 32-1100B
preamplifier connected to a 486-based computer with a Turtle Beach
Tahiti sound card.  Sound was digitized at a sample rate of 11,000
samples per second in 16-bit linear pulse code modulation (PCM) for
all of our pulse counting runs.  Preamplifier and sound card gains
were constant for all of our recordings, and all crumples were
performed at a distance of 12'' from the microphone.

	Reference recordings taken at a sample rate of 44,000 samples
per second and 16-bit linear PCM of crumpling demonstrated that the
power spectrum for the crumpling of Xerox 4024 paper is peaked around
2 kHz, below the 5.5 kHz Nyquist frequency set by our usual sampling
rate.  Similar signals observed in magnetic \cite{barkA} \cite{barkB}
and martensitic \cite{microfrac} systems exhibit a broad range of
frequencies, due to either a broad range of pulse durations and shapes
\cite{powspec} or on the time correlations between events \cite{microfrac}.
Our pulses have much less structure. (\ref{fig:short}) We finds that
large events are impulsive and the relationship between duration and
energy is consistent with predominantly exponential decay, and we do
not observe nontrivial scaling in the power spectrum.  (see Figure
\ref{fig:duration})

	To remove the DC offset from our data, we measured the median
of the amplitude and subtracted it.  We then integrated the energy in
bins of fixed duration and compared the energy in each bin to a
threshold.  Contiguous runs of bins over threshold were considered to
be single pulses and the pulse end time, duration, energy and peak
amplitude were written into a data file.  We then plotted histograms
using bins logarithmically spaced over pulse energy; error bars are
${\pm} 1 \sigma$ assuming Poisson statistics.  Figure \ref{fig:short}
illustrates the process by which two pulses are identified.  The RMS
amplitude of noise in the anechoic chamber with the human crumpler
sitting motionless inside was 27.5 in computer units.

	Our pulse counting algorithm has two arbitrary parameters, the
bin duration and the amplitude threshold and we found it important to
choose them wisely; because the parameters are arbitrary, we would
expect our histogram to be insensitive to moderate changes in the
parameters (of order 50 \%) when pulses are being accurately counted .
When we chose a bin length much shorter than 1 ms, our oscillating
signal would drop below the threshold prematurely and our algorithm
would inappropriately fragment the pulses; in some of our early plots
made before we started binning (when our bin size was effectively one
sample) we observed false power laws spanning up to six decades in
energy due to this.  For our early analysis, influenced by
\cite{barkB}, we set our threshold to the median of bin power but we
found with some data sets the histograms were strongly influenced by
small changes in the threshold.  Investigating this, we discovered
that when our threshold was low, long (duration $> 50 {\rm ms}$)
bursts of low amplitude noise caused presumably by paper friction or
some mechanism other than of interest were causing clearly separate
events to be merged.  We found that the severity of this would vary
depending on the method and speed of crumpling, since slower crumpling
would spread the pulses out in time making them easier to separate and
because some of the data sets, such as strong cylindrical crumpling of
drawing paper, had much more unwanted paper noise than other sets,
such as weak crumpling of paper with a grid.

	We searched for a set of parameters that would accurately
isolate pulses for all of our data sets and we converged on a bin
length of 30 samples (2.7 ms) and a threshold amplitude of 50 computer
units.  (The threshold energy equals the threshold amplitude squared
times the bin length) We tested the pulse identification algorithm in
two ways. (1) The output of the pulse counter was verified by
comparing a sample of the pulses counted to a manual analysis of the
set.  Pulses identified by the algorithm were examined by eye to
determine if they actually were impulsive events (in contrast to
extended noise bursts) and to determine if they were inappropriately
split or merged.  We considered the output of the algorithm acceptable
when 90\% or more of the pulses in an energy bin were correctly
identified.  In addition, we checked the accuracy of integration for
the weak crumpling sets (the sets of best quality) and it was found
that our pulse counter with standard settings consistently
underestimated the energy of pulses by $730 \pm 260 \sigma$ in
arbitrary units independent of pulse energy from smallest to largest.
This is what is expected, given our algorithm, since our threshold
should cut off an exponential tail of nearly constant area.  We
estimated the cutoff energy below which identification errors were
unacceptable for at least one set in each category.  (2) We then
developed a faster alternative test of pulse identification in which
we would make pulse energy histograms increasing and decreasing the
pulse threshold by 50\%.  Near the cutoff energy determined by the
manual test the curve would secularly veer out of the error bars.  We
chose this as a criterion for setting the lower bounds on our
histograms.  One weak crumpling set, (when we weak crumpled an
initially flat sheet) had significant merging problems up to
$E=20,000$ because the sheet was crumpled much more rapidly than later
experiments.  In our other weak crumpling sets, pulse identification
was accurate down to $E=1,000$.  In our strong crumpling sets we have
problems with merging and spoofing below $E=1,000$ to $E=10,000$
depending on the set.  We believe that with a lower threshold and
shorter bin size we can accurately count pulses with lower energies in
most of the weak crumpling recordings, but we chose to use a
consistent set of parameters for all of our sets.  Power law behavior
appears to continue for another decade in our triangular grid/weak
crumpling experiments with less conservative parameters.
\footnote{visit URL
http:www.msc.cornell.edu/~houle/crumpling/}

\begin{figure}
\centerline{\psfig{figure=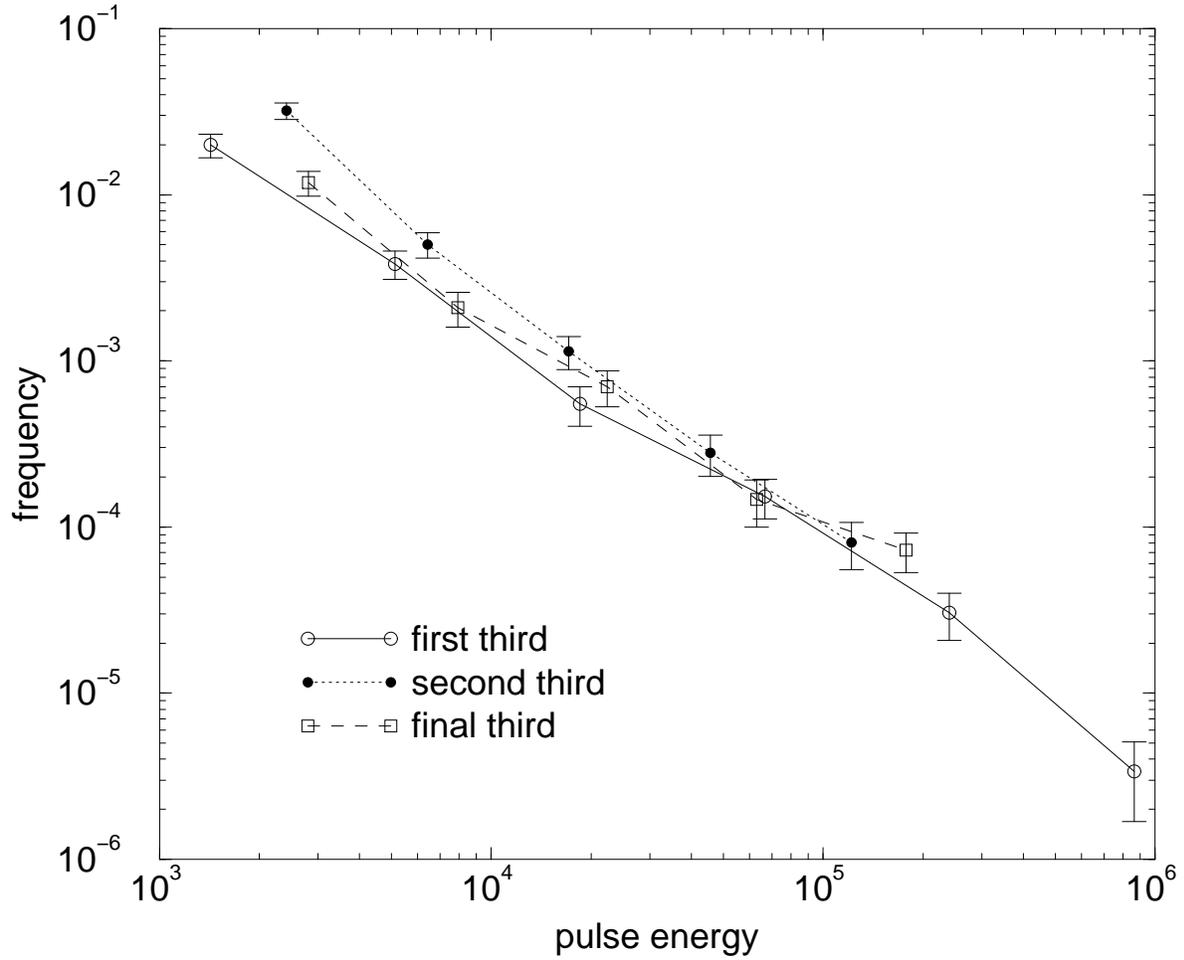,width=7.0truein}}
\caption{Strong cylindrical crumples:  time evolution.  This
is a sum
of three runs performed with Xerox 4024 paper.  The time
series were divided
in thirds over time.  Although it appears that some 
very energetic events
occur in the early stages of crumpling, there appears
to be no systematic
variation at other energies.}
\label{fig:time}
\end{figure}

        To search for time dependence in the energy distribution of
sound pulses produced by strong crumpling we performed three crumples
using the hand and cylinder methods with respectively letter size (8.5
x 11'') and 8.5'' square Xerox 4024 paper.  We subdivided the sets
over time into thirds and combined the crumples to improve statistics.
Figure \ref{fig:time} shows the result for cylindrical crumpling.
About five exceptionally large events, spread out between the three
crumples, cause the histogram for the first third in time to extend
for a decade further than the others.  There seems to be no systematic
difference between the last two thirds of the crumpling process, or in
the distribution of pulses of low to moderate energy.  We made a
similar graph for strong crumpling by hand that displayed even less
evidence for time variation; hand crumpling did not produce
exceptionally large events in early crumpling nor any systematic
variation in the pulse energy distribution.

\begin{figure}
\centerline{\psfig{figure=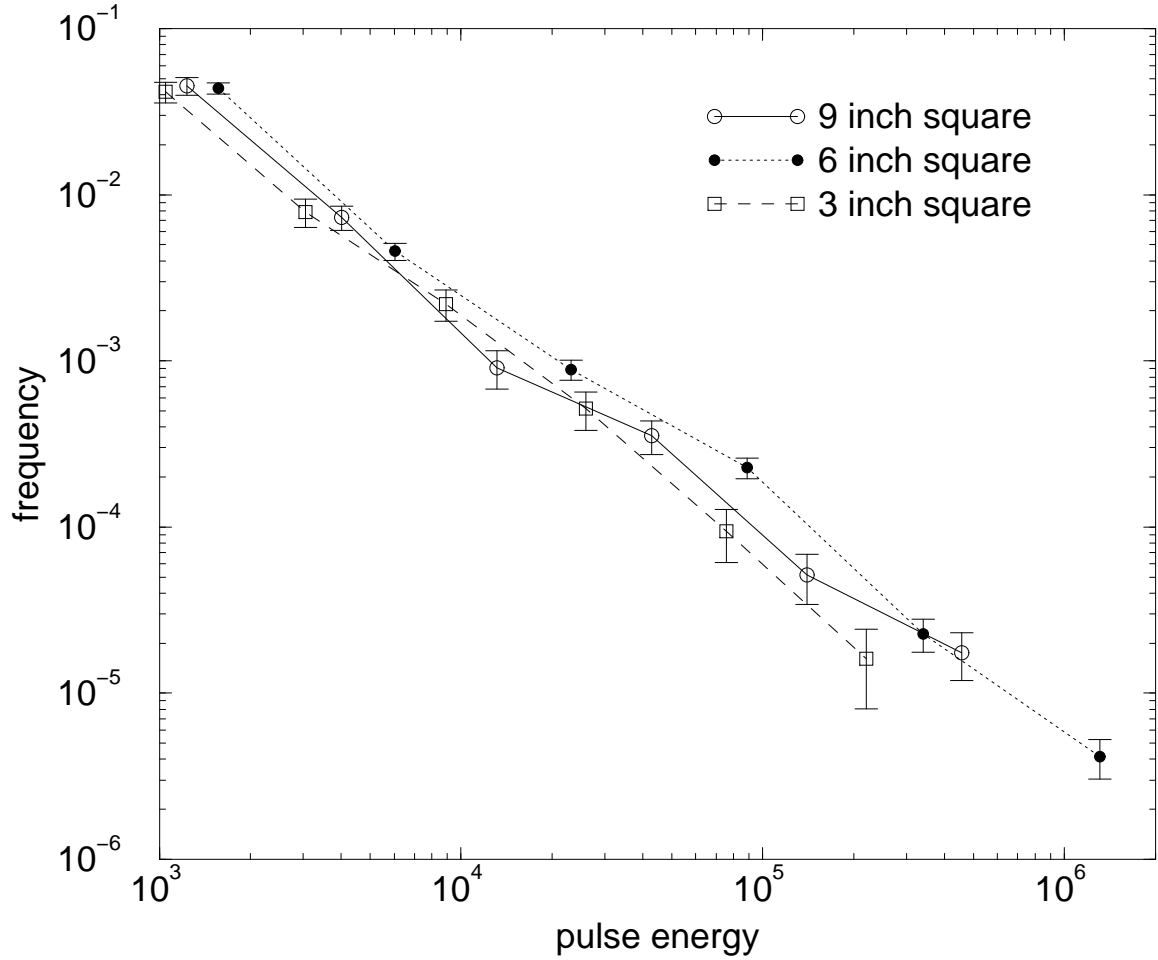,width=7.0truein}}
\caption{Strong cylindrical crumpling:  size variation.  Medium 
weight
drawing paper was cut into squares and crumpled 
using the cylindrical
method.  Larger numbers of smaller squares were 
crumpled to combat the
loss of events,  and the vertical axis is 
normalized over cumulative
crumpled sheet area.}
\label{fig:size}
\end{figure}

        To study finite size effects in paper crumpling, we performed
sets of strong cylindrical crumples were with square sheets of medium
drawing paper (Carolina Pad Company item 54115) of sides 9", 6" and 3"
and the cylinder diameter one third the side of the paper.  Drawing
paper is considerably thicker than Xerox 4024 paper and presumably
will have a longer characteristic length scale.  A single sheet of 9"
square paper was crumpled, four sheets of $6'' \times 6"$ and nine
sheets of $9'' \times 9"$.  The vertical axis of the histogram in
figure \ref{fig:size} is normalized to sheet area.  Since the sheet
can only fragment into smaller facets with the passing of time, the
natural assumption that pulse energy is determined primarily by facet
size is contradicted by the lack of both size and time dependence.

\begin{figure}
\centerline{\psfig{figure=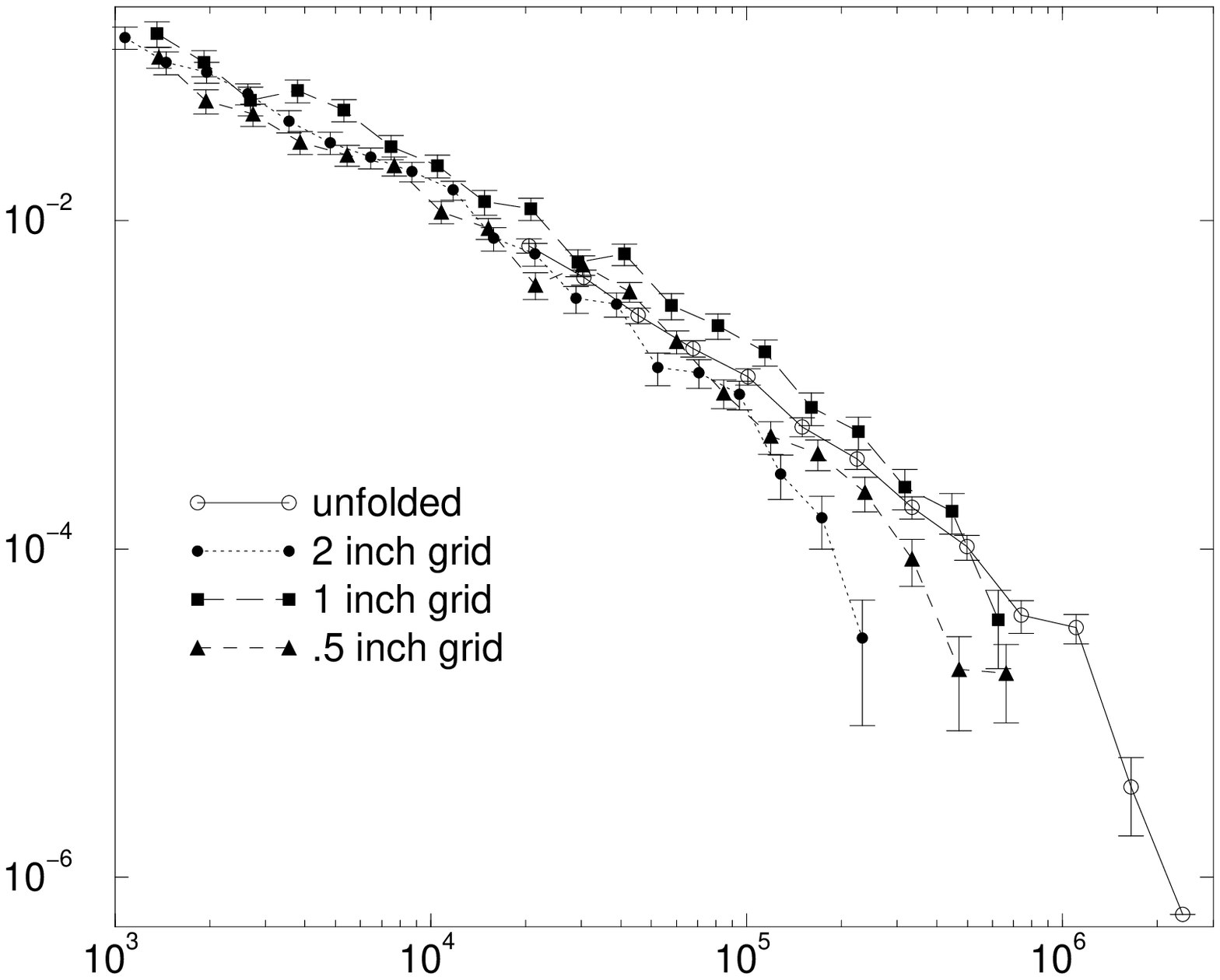,width=7.0truein}}
\caption{Weak crumpling of triangularly gridded Xerox 
4024 paper,  normalized
to fifty cycles.  No systematic dependence of pulse
energy distribution
on grid scale is seen. }
\label{fig:triangles}
\end{figure}

	Because we were interested in isolating the effect of existing
creases from that of self avoidance, which would surely be important
in a dense ball, we made recordings of the weak crumpling of
pre-creased and crumpled sheets using Xerox 4024 paper on 3'' diameter
cylinders.  These sets were of excellent quality, since pulses were
well separated in time ($\gg 100 {\rm ms}$), noise from paper friction
was almost completely eliminated, and the number of pulses counted was
much greater than the other experiments.  We weak-crumpled an
uncreased sheet, a sheet of previously hand crumpled paper, and a
sheet of previously cylinderically crumpled paper.  We also
weak-crumpled sheets that had been hand-creased along triangular grids
with interline spacings of 2", 1.5", 1.0", 0.75" and 0.50".  Figure
\ref{fig:triangles} shows that the introduction of a creased grid clearly
suppresses large events but shows no systematic relationship between
the grid spacing and the energy scale at which suppression occurs.  It
proved possible to collapse the probability distributions for the
various triangular grid spacings and the previously cylindrically
crumpled grid by multiplying the energy and probability densities by
constants, but the constants required appear to be random, showing no
secular dependence on the grid size.  Comparing early and late parts
of weak crumpling runs involving up to 100 cycles we found no evidence
for time dependence.

\begin{figure}
\centerline{\psfig{figure=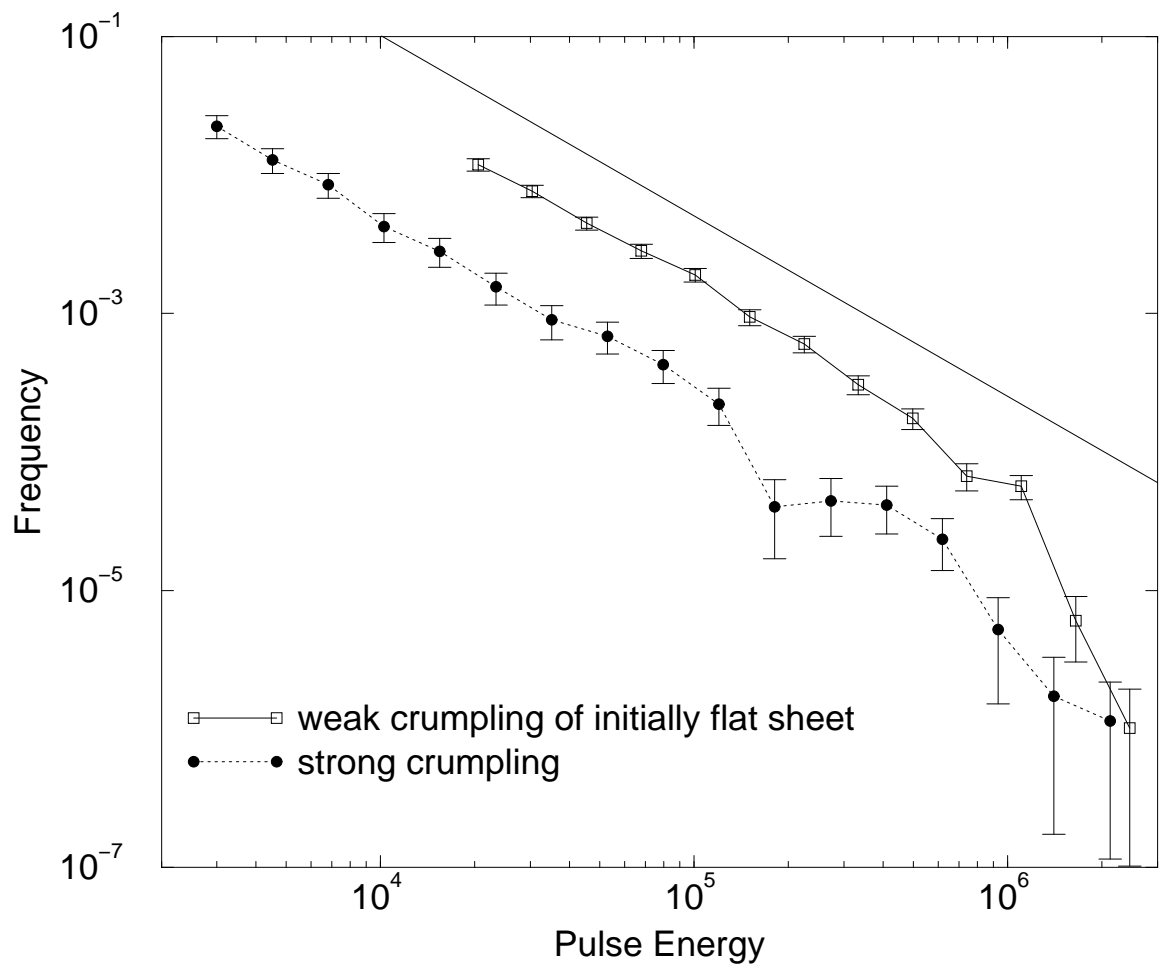,width=7.0truein}}
\caption{Strong vs. weak cylindrical crumpling:  Xerox 4024 
paper,  sum
of three runs of strong cylindrical crumpling is compared
to 80 repetitive
cycles of weak cylindrical crumpling.  Overdrawn line 
has a slope of -1.3}
\label{fig:cyl}
\end{figure}

\begin{figure}
\centerline{\psfig{figure=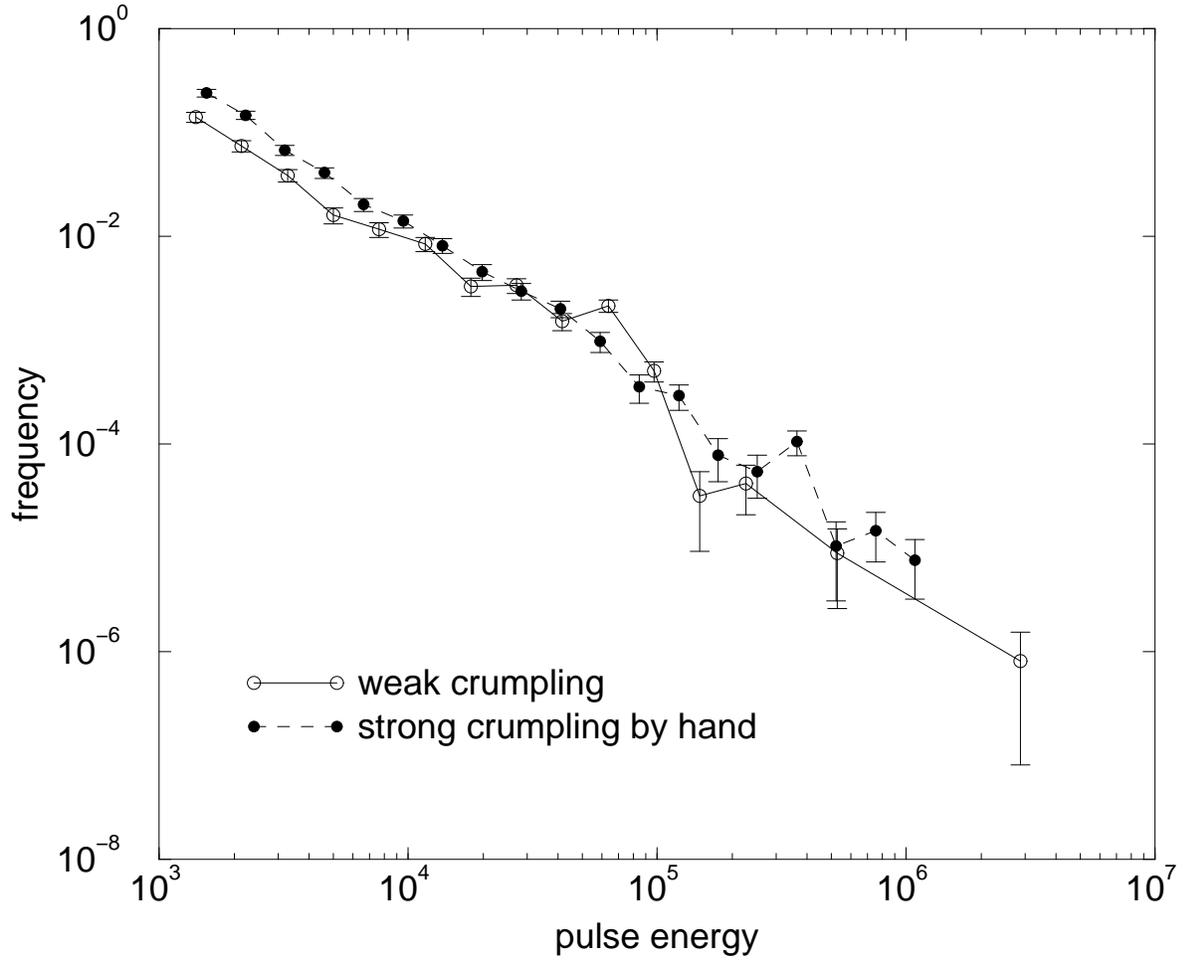,width=7.0truein}}
\caption{Weak crumpling of a previously hand crumpled
sheet compared to
the sum of three strong crumplings by hand.  In both
cases we observe
that larger events are suppressed more strongly
than in cylindrical
crumpling.}
\label{fig:hand}
\end{figure}

	Figure \ref{fig:cyl} compares weak cylindrical crumpling and
strong cylinderical of an initially flat sheet.  Since many other
systems produce pulses with a power law distribution in energy
\cite{barkA} \cite{barkB} \cite{microfrac} and it appears that the
histograms could be well-fit by a line on a log-log plot, we fit a
power law of the form $ p(E) = E^\alpha$ to our histograms.  Over the
energy range $E = 10^4 - 10^6$ we get $\alpha= -1.30 \pm .04$ for
strong crumpling and $\alpha=-1.30 \pm .03$ for weak crumpling.  We
then combined all of the finite size runs using medium paper since we
saw no dependence on size and fit an exponent of $\alpha = -1.32 \pm
.03$ over the range $E=10^3 - 5 \times 10^5$, which is compatible with
the histogram from the 9'' sheet alone with $\alpha=-1.24 \pm .06$.
Larger events appear to be suppressed more strongly when a sheet is
strongly crumpled by hand( $\alpha = -1.59 \pm .09$ over the range of
the plot), and when a previously hand crumpled sheet is weakly
crumpled on cylinders ($\alpha = -1.59 \pm .04$),
Figure~\ref{fig:hand}.  We believe that the statistical errors in the
fit exponents are much smaller than the systematic errors.  The
observed difference between strong hand crumpling of virgin paper and
weak cylindrical crumpling of pre-crumpled paper is statistically
significant. (Figure \ref{fig:hand})

	Our data is compatible with the assertion that the energy
released when a facet buckles is insensitive to the size of the facet.
Although it is possible that we are not probing a small enough length,
we see no systematic dependence on facet size when we introduce a
grid.  In addition, since facets are formed by the fragmentation of
larger facets, the size scale of facets on the sheet can only decrease
over time; the lack of time dependence suggests a lack of size
dependence.  If we presume that a nearly constant fraction of the
elastic energy difference between the buckling metastable and final
states is converted into sound, the pulse energies may be reflective
of the distribution of the elastic energy stored in and around the
facets.  If we vary the length scale of an elastic sheet with a
constant shape, the energy of bending scales as $L^0$ and the energy
of stretching scales as $L^2$ where $L$ is the length.  If the energy
were primarily stored in bending, the energy stored in a facet will
have no direct dependence on the area of the facet.  However, it has
been proposed that when the configuration of an elastic sheet
minimizes the sum of bending and stretching energies, deformation can
isolate itself in temporary ridges (a purely elastic phenomenon
distinct from the permanent creases) with energy scaling as $L^{1/3}$.
\cite{ridgesA} \cite{ridgesB}  If the energy emitted during the shift
between two stable configurations scaled as weakly as $L^{1/3}$ this
could explain our lack of observed finite size dependence; this is
plausible if the surface can be understood as an interacting network
of ridges as considered in \cite{ridgesB}.  It is possible that we
observe a small number of very large events only in the earliest
stages of crumpling and in the weak crumpling of an initially flat
sheet because the existence of an extensive crease network in other
situations might limit the range of facet {\it shapes}.  Whereas a
flat or nearly flat sheet forms very sharp cones when stress is
applied at the edges (try it), a sheet with a crease network is likely
to deform by bending at the creases instead, suppressing facet
configurations that may produce high energy events.

	The fact that the oscillation frequency and decay time of
ring-downs depends on the type of paper and appears to be the same
with both the standard 11 kHz sample rate used for standard recordings
and the 44 kHz sample rate used for reference recordings indicates
that the ring-downs are a property more of the paper than of the
recording system.  However, it is interesting that oscillation
frequency of pulses does not depend strongly on the pulse energy or
the degree of crumpling of the sheet.  A possible explanation is that
the buckling of a facet concentrates energy into a small area.  Such a
process would halt at a length scale set by the thickness of the
paper, disturbing the surface with a wavenumber insensitive to facet
size and hence little variation in the frequency of oscillation.

	In our experiments we have found that the crumpling of paper
generates acoustic pulses with a distribution in energy that varies
nonexponentially over at least three orders of magnitude and
compatible with power law scaling over at least two.  We also find
that the pulse distribution appears to vary little over time or change
in the length scale.  Our use of a cylindrical geometry for strong and
weak crumpling makes it possible to crumple paper by a process that is
both mathematically and practically well defined, providing a handle
for mechanization and theory.  However, we do find that cylindrical
crumpling may produce a different experimental pulse energy
distribution than hand crumpling, perhaps because cylindrical
crumpling is fundamentally anisotropic and produces a less compact
object than hand crumpling.

This project was supported by DOE grant DE-FG02-88-ER45364 and NSF
grant DMR-9419506 .  We thank Wolfgang Sachse for allowing us access
to an anechoic chamber, Naresh Kannan for logistic support and many
good discussions, as well as helpful discussions with Karin Dahmen,
Olga Perkovi{\'c} and Eric Kramer.  More information about this
research, including audio samples of crumpling paper can be found on
the World Wide Web, URL
http://www.msc.cornell.edu/~houle/crumpling/.

\eject

\eject
\begin{figure}
\centerline{\psfig{figure=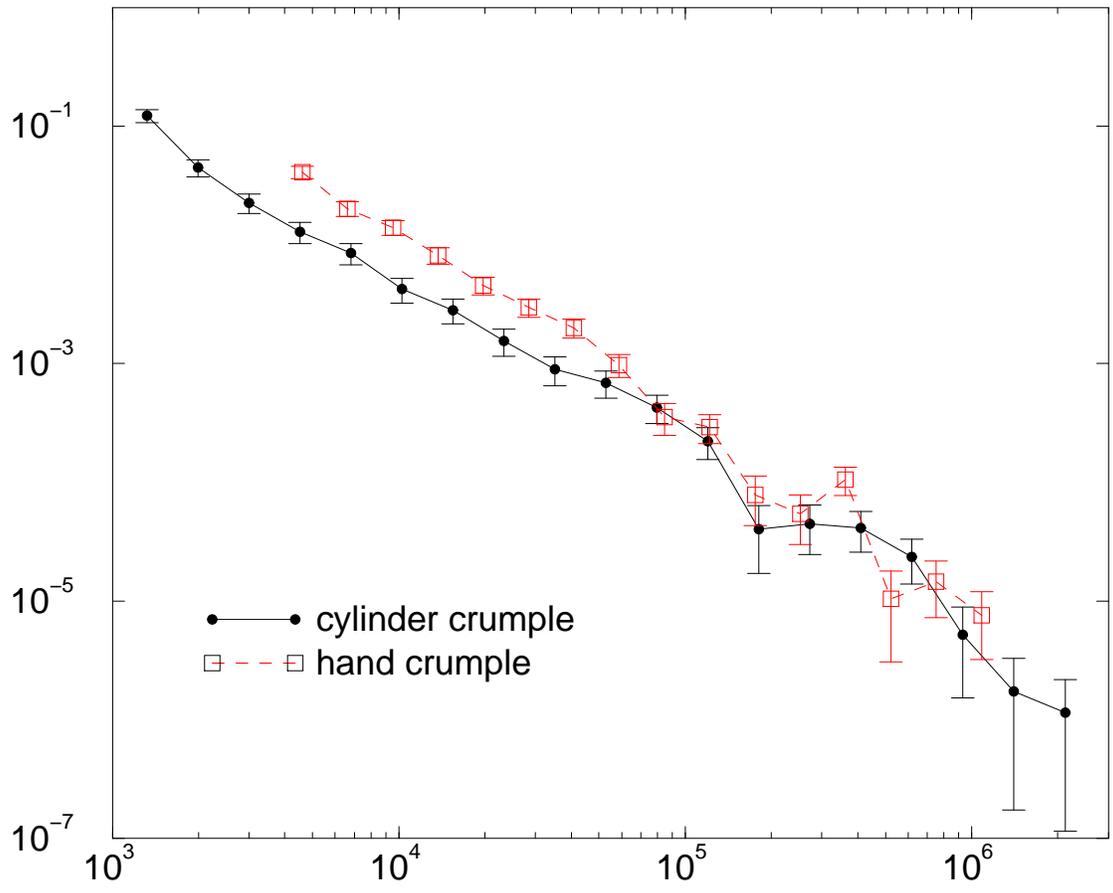,width=7.0truein}}
\caption{Strong cylindrical crumples and
 hand crumples:  Xerox 4024 duplicator paper was
crumpled strongly in hand
and using the cylindrical method.  In both cases a sum
of three crumpling
runs is shown.  Error bars are $\pm 1 \sigma$ predicted
by Poisson statistics.}
\label{fig:strong}
\end{figure}

\eject

\eject

\eject

\eject

\eject

\end{document}